\documentclass[journal,twoside]{IEEEtran}
\ifCLASSINFOpdf
\else
\fi
\usepackage[english]{babel}
\usepackage{amsmath,graphicx}

\usepackage{filecontents}
\usepackage[font=footnotesize]{caption}
\usepackage[latin1]{inputenc}  
\usepackage[T1]{fontenc}
\usepackage{graphicx}          
\usepackage{psfrag}           
\usepackage{amssymb}           
\usepackage{latexsym}
\usepackage{amsmath}
\usepackage{mathtools}
\usepackage{tikz}
\usetikzlibrary{shapes,arrows}
\usepackage{verbatim}
\usepackage{tabularx}
\usepackage{color}
\usepackage{algorithm}
\usepackage{makecell}
\usepackage{algpseudocode}
\usepackage{algorithmicx}
\usepackage{algpseudocode}
\usepackage{booktabs}
\usepackage{pdflscape}
\usepackage{psfrag}
\usepackage{lipsum,multicol}
\usepackage{latexsym}
\usepackage{tikz}
\usetikzlibrary{shapes,arrows}
\usepackage{epsfig}
\usepackage{color}
\usepackage[numbers, square, comma, sort&compress]{natbib}
\usepackage{hyperref}

\makeatletter
\renewcommand{\ALG@beginalgorithmic}{\small}

\newcommand{\SNR}{{\mbox{SNR}}}
\newcommand{\MSE}{\mbox{MSE}}
\newcommand{\erf}{\mbox{erf}}
\def\NoNumber#1{{\def\alglinenumber##1{}\State #1}\addtocounter{ALG@line}{-1}}
\algdef{SE}[DOWHILE]{Do}{doWhile}{\algorithmicdo}[1]{\algorithmicwhile\ #1}%
\usepackage{babel}
\makeatother

\begin{document}


\title{Phase-Noise Mitigation in OFDM by Best Match Trajectories}








\author{~Senay~Negusse,~Per~Zetterberg,~Peter~H{\"a}ndel
\thanks{S. Negusse (negusse@kth.se), P. Zetterberg (perz@kth.se), and P. H{\"a}ndel (ph@kth.se),  
are with the Department of Signal Processing, ACCESS Linnaeus Center, KTH Royal Institute of Technology, 
100 44 Stockholm, Sweden. }}


%

\markboth{}%
{}

\maketitle












\begin{abstract}
This paper proposes a novel approach to phase-noise compensation. The basic idea is to approximate the phase-noise statistics by a finite number of realizations, i.e., a phase-noise codebook.  The receiver then uses an augmented received signal model, where the codebook index is estimated along with other parameters.  The realization of the basic idea depends on the details of the air interface, the phase-noise statistics,  the propagation scenario and the computational constraints. In this paper, we will focus on a MQAM-OFDM system with pilot sub-carriers  within each OFDM symbol. The channel is frequency selective, fading and unknown. A decision-feedback method is employed to further enhance performance of the system. Simulation results are shown for uncoded and coded systems to illustrate the performance of the algorithm, which is also compared with previously employed methods. Our simulations show that for a 16-QAM coded OFDM system over a frequency selective Rayleigh fading channel affected by phase noise with root-mean-square (RMS) of 14.4 degrees per OFDM symbol, the proposed algorithm is 1.5dB from the ideal phase-noise free case at a BER of $10^{-4}$. The performance of the best reference scheme is 2.5dB from the ideal case at BER of $10^{-4}$. The proposed scheme is also computationally attractive.
\end{abstract}
\begin{keywords} OFDM, phase-noise, BER and SNRs, codebook, fading channel, channel estimation, pilot sub-carrier.
\end{keywords}
\IEEEpeerreviewmaketitle
\section{Introduction\label{sec:1}}
Orthogonal frequency division multiplexing (OFDM) is adopted by most of the current and future telecommunication standards for high-rate data transmission, particularly in wireless communication systems. Its resilience to multipath channel fading, the spectral efficiency it provides as well as the simplicity of the equalization, has enabled OFDM to remain the most popular modulation scheme. However, OFDM is known to be sensitive to various hardware imperfections, the Dirty-RF effect \cite{Fettweis,Schenk}, originating in the transceiver hardware. \setlength{\parskip}{0pt}
\subsection{Previous Works}
The phase-noise (PHN) phenomenon and its underlying effects on various OFDM systems were studied extensively in \cite{Schenk,Demir, Armada,Petrovic,Ville,Sahai,Pitarokoilis}. The effect of PHN on OFDM systems is classified into two components. The common phase error (CPE), which rotates all the sub-carriers in one OFDM symbol by a common phase distortion, and the inter-carrier interference (ICI), which arises due to the loss of orthogonality between each sub-carrier. Frequency domain approaches to PHN estimation and compensation  mainly deal with the CPE and ICI components separately, while time domain approaches attempt to compensate for both jointly. It has been shown that a significant improvement in performance can already be achieved with CPE correction only by treating the ICI term as an additive Gaussian noise \cite{Petrovicc, Sridharan, Nie}. However, CPE correction only is not always sufficient for high rate transmission, therefore ICI compensation is  necessary. 

Most ICI compensation techniques employ decision-directed feedback (DD-FB) for frequency domain PHN estimation \cite{Petrovic, Bittner,Munier,Corvaja,Yue,Khanzadi}. A vast majority of works assume a known channel frequency response \cite{Petrovic,Bittner,Yue, Khanzadi,Syriala,Tchamov, Linn}. For unknown channel, joint channel and PHN estimation is considered in \cite{Munier} where an ICI reduction scheme over a Rayleigh fading channel is presented in which the PHN process within an OFDM symbol is modeled as a power series. Although the method presented showed a significant bit error rate (BER) improvement, the cost in computation is large. A less complex method is presented in \citep{Corvaja} where the estimate of channel and CPE at pilot subcarriers are interpolated to obtain channel frequency response and CPE followed by a DD-FB loop to estimate ICI components.   

A non-iterative compensation scheme in which the CPE between consecutive OFDM symbols are interpolated linearly to estimate the time varying PHN is presented in \cite{Syriala,Tchamov}. In \cite{Chun-Ying}, a method of suppressing the ICI is presented by linearly combining the cyclic-prefix and the corresponding OFDM samples. The linear coefficients are obtained such that the ICI power is minimized. Other methods using the Bayesian framework for joint channel and PHN estimation have been proposed in  \cite{Khanzadi2,Septier,Simoens,negusse}. Soft-input maximum a posteriori and extended Kalman smoother are proposed in \citep{Khanzadi2}. In \cite{Septier,Simoens,negusse},  Monte-Carlo methods are employed to approximate the posterior probability distribution of the unknown quantities for PHN tracking and channel estimation. Although they are shown to provide good BER performance, Monte-Carlo methods, such as particle filters, are known for their numerical complexity. The main interest in the paper lies on PHN compensation over unknown channels using pilot sub-carriers. However, the proposed method also uses  DD-FB loop for improved performance. For the known channel the proposed method is benchmarked against \cite{Petrovic}. For known and unknown frequency selective Rayleigh fading channels the methods in \cite{Munier,Corvaja} are also used as benchmarks. The problem setting in which the methods presented in \cite{Petrovic}, \cite{Munier,Corvaja} and  are employed is identical to the setting the proposed algorithm.

\subsection{Contributions}
In this paper, we employ a simple and novel scheme for PHN cancellation which  is shown to have superior performance in terms of BER compared to previously proposed methods. The proposed technique can be applied to any assumed distribution of the PHN process, $\theta(n)$. Additionally, no approximation based on the magnitude of the PHN is taken into account, e.g. $e^{j\theta(n)}\simeq 1 + j\sin(\theta(n))$, as is common in some previous works. 

A codebook of $K$ vectors representing a set of trajectories which aim to closely match with the PHN realization is used.  An uncountable set of possible PHN realizations is represented by a codebook with $K$ quantized trajectories, which are stored at the receiver.\footnote{The technique is similar to the design of vector quantization codebook where a set of $n$ vectors from some $m$-dimensional space is efficiently represented by a codebook with a smaller set of vectors from $m$-dimensional space \cite{Gersho}.}  The complexity of the algorithm is therefore dependent on the number of trajectories in the codebook, $K$. For a moderately small sized codebook, it is shown that the impact of the PHN is significantly reduced (e.g. with $K=27$, there is 60\% reduction in terms of the effective PHN mean square error (MSE), see Section \ref{MSE}). 

In this paper, we focus on a solution where the trajectory that minimizes the Euclidean distance between the constellation of the received symbols at the pilot positions and the constellation of the known pilot symbols is chosen. Additionally, a DD-FB technique is employed such that both the estimate of data symbols from the channel decoder and pilot symbols are used to compute the Euclidean distance. The proposed technique can also be used for combined channel estimation and PHN compensation schemes, in which MMSE based channel estimation is employed for each of the vectors in the codebook. The channel estimate corresponding to the trajectory in the codebook which best approximates the PHN realization is chosen. We will show by the simulation that, for AWGN and fading channels with a known frequency response, the proposed scheme outperforms previously proposed PHN compensation techniques. In this scenario, the proposed method  does not employ DD-FB while the reference schemes require previously detected symbols for ICI estimation. Simulation results are also shown for unknown channel frequency response, in which the proposed method employs a DD-FB technique for combined channel estimation and PHN compensation. It is shown that the method presented provides improved performance compared to previously proposed methods. Additional results based on a PHN process modelled as  Ornstein-Steinbeck is also shown to illustrate the applicability of the proposed method for a PHN model other than Weiner process.

The paper is organized as follows. Sec. \ref{sec:2} introduces the system model and parameters while Sec. \ref{sec:3} presents design of the codebook for Weiner PHN along with some performance analysis. In Sec. \ref{sec:4}, details of implementation of the codebook
for PHN compensation together with channel estimation. Simulation results are shown in Sec. \ref{sec:5} and a computational analysis of the proposed algorithm as well as some of the reference methods is given in Sec. \ref{sec:6}. We end with concluding remarks in \ref{sec:7}.

\section{System Model and Proposed Approach\label{sec:2}}
We consider an OFDM transmission system over a fading time varying channel with $L$ multipath taps employing QAM modulation technique. The OFDM system is assumed to have FFT size of $N$ of which $N_p$ are pilot subcarriers and with a cyclic prefix of length $N_{cp}$ such that $N_{cp} \geq L$. To obtain the $m$-th OFDM symbol of duration $T$, a stream of data bits is divided to $N-N_p$ groups of $M$-bits which are mapped to $2^M$-QAM complex symbols, $S_m(k)$.  Assuming perfect frequency and timing synchronization, the baseband representation of the received time domain signal, $y_m(n)$, is given by
\begin{equation}\label{eq1}
y_m(n)=e^{j\theta_m(n)}\bigg(\sum_{\ell=0}^{L-1}h_m(\ell)s_m(n-\ell)\bigg) + w_m(n)
\end{equation}
where $s_m(n)$ is the $N$ point inverse FFT (IFFT) of the complex transmitted symbol, $S_m(k)$, $h_m(i)$ is the fading complex channel pulse response with $L$ propagation paths, $w_m(n)$ is a zero mean complex circular Gaussian channel noise with variance $\sigma^2_w$; and $\theta_m(n)$ is the PHN sample at the receiver at time index $n$ of the $m$-th OFDM symbol. The received signal is sampled at a frequency $f_s = (N+N_{cp})/T$, where $T$ is the OFDM symbol duration. 

\begin{figure*}[t]
\centering
\includegraphics[scale=0.63]{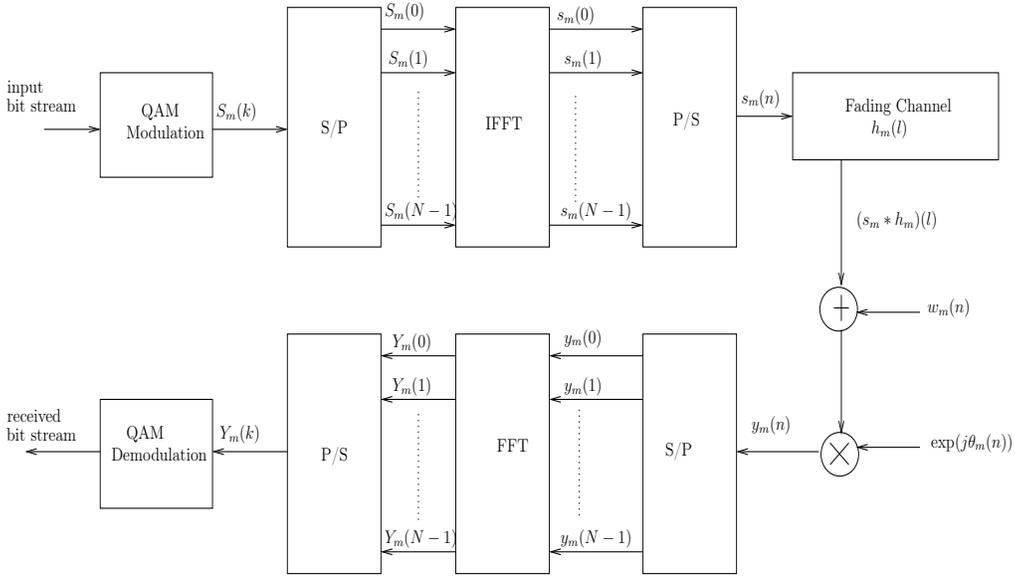}
\caption{Block diagram of the OFDM system.}
\label{fig1_a}
\end{figure*}

After removing the cyclic prefix, an FFT operation is applied to the received signal such that the $k$-th subcarrier is given by
\begin{equation}\label{eq2}
\begin{split}
Y_m(k)&=\frac{1}{\sqrt{N}}\sum_{n=0}^{N-1}\sum_{\ell=0}^{L-1}h_m(\ell)s_m(n-\ell)e^{j(\theta_m(n)-2\pi k n)}\\
 & + \frac{1}{\sqrt{N}}\sum_{n=0}^{N-1}w_m(n)e^{-j2\pi k n}\\
&= S_m(k)H_m(k)\underbrace{A_m(0)}_{\displaystyle  \mbox{CPE}} +  \underbrace{\sum_{\substack{l=0 \\
l\neq k}}^{N-1} S_m(l)H_m(l)A_m(l-k)}_{\displaystyle  \mbox{ICI}} \\ &+ W_m(k)
\end{split}
\end{equation}
where $A_m(k)$ and $W_m(k)$ are respectively the FFT of $e^{j\theta_m(n)}$ and that of the additive complex Gaussian channel noise $w(n)$. $H_m(k)$ is the instantaneous frequency response of the  multi-path channel which is assumed to be constant within one OFDM symbol. Figure \ref{fig1_a} shows the block diagram of the system model. Equation \eqref{eq2} can further be written in matrix form as 
\begin{equation}\label{eq3}
\mathbf{Y}_m= \mathbf{A}_m \mathbf{S}^d_m \mathbf{H}_m + \mathbf{W}_m
\end{equation}
where $\mathbf{S}^d_m=\mbox{diag}(S_m(0),..,S_m(N-1))$ is an $N \times N$ matrix representing diagonalization of the symbol vector $\mathbf{S}_m$; and the $N \times N$ matrix $\mathbf{A}_m$ is a unitary circulant matrix containing DFT coefficients of $e^{j\theta_m(n)}$ such that $\mathbf{A}_m(k,l)=A_m(l-k)$. $\mathbf{H}_m$ and $\mathbf{W}_m$ are vectors of the channel frequency response and the complex Gaussian noise respectively. Estimation of the realization of the PHN process, $\mathbf{\theta}_m=[\theta_m(0), \cdots \theta_m(N-1) ]^T$,  would enable compensation of CPE as well as ICI. 
\subsection{Approximate Model}
We propose a method where the PHN realization at the $m$-th OFDM symbol is approximated by one of a set of $K$ vectors,  $\{\mathbf{\phi}^k\}_{k=1}^K$, where $\mathbf{\phi}^k =[\phi_k(0), \phi_k(1), \cdots, \phi_k(N-1)]^T$ plus an initial phase offset, $\psi_m \in [0, 2\pi)$. Associated with each vector, $\mathbf{\phi}_k$, is a corresponding unitary circulant matrix containing DFT coefficients of $\{e^{j\phi_k(l)}\}_{l=0}^{N-1}$ denoted by $\mathbf{\widehat{A}}$. The signal model in \eqref{eq3} is then modified as 
\begin{equation}\label{eq4}
\mathbf{\bar{Y}}_m=e^{j\psi_m}\mathbf{\widehat{A}}_{k_m}\mathbf{S}^d_m \mathbf{H}_m + {\mathbf{W}}_m
\end{equation}  
which is an approximation of \eqref{eq3}, where the PHN DFT coefficient matrix, $\mathbf{A}_m$, is now replaced by  $e^{j\psi_m}\mathbf{\widehat{A}}_{k_m} $, which is a function of an index, $k_m$, that takes on a value between $1$ and $K$ with equal probability. The framework can be extended to unequal probabilities but this is left for future work. The random variable $\psi_m$, which is assumed independent of $k_m$, is uniformly distributed and models the initial phase of the symbol. Both $k_m$ and $\psi_m$ are independent from symbol to symbol. The design of the codebook $\{\phi^k\}_{k=1}^K$ is yet to be discussed in a later section. It should be noted that this approximate model is only used for the purpose of deriving the algorithm. Data from the original model in \eqref{eq3} is used when testing and validating the proposed algorithm by simulation.

\subsection{Cost Function}
Any estimator derived in the absence of PHN can be written as  
\begin{equation}\label{eq5}
\widehat{\Omega} = \arg\min_{\Omega}\mathcal{C}(\mathbf{\bar{Y}}_1, \dots, \mathbf{\bar{Y}}_q),
\end{equation}
where $\widehat{\Omega}$ is the estimated parameter, $\mathcal{C}$ is some form of optimization criterion and $q$ is the number of OFDM symbols. It should be noted that $\mathbf{\widehat{A}}_{k_m}$ and its inverse $\mathbf{\widehat{A}}_{k_m}^{-1}$ are unitary circulant matrices. As a result, $\mathbf{\widehat{A}}_{k_m}$ is invertible and $\mathbf{\widehat{A}}_{k_m}^{-1}\mathbf{W}_{m}$ is statistically identical to $\mathbf{W}_{m}$. We may therefore generalize the criterion \eqref{eq5} as 
\begin{equation}\label{eq6}
\widehat{\Omega} = \arg\min_{\Omega, k_1,\cdots,k_q,\psi_1,\ldots,\psi_q}\mathcal{C}(e^{-j\psi_1}\mathbf{\widehat{A}}_{k_1}^{-1}\mathbf{\bar{Y}}_1, \dots, e^{-j\psi_q}\mathbf{\widehat{A}}_{k_q}^{-1}\mathbf{\bar{Y}}_{q}).
\end{equation}
for the model \eqref{eq4}. This includes, for instance, maximum-likelihood estimators where now the estimator in \eqref{eq6} is also the maximum likelihood estimator for the model \eqref{eq4}. An example of a use of the proposed new estimator would be in an OFDM system with an equalizer and Viterbi decoder and a pilot subcarrier. In such a system, the optimization function could be the sum path metric. In this case, \eqref{eq6} implies that the entire receiver is run for all possible hypotheses of $k_1 \dots k_q$ (of which there are $K^q$), and selects the one with the lowest path metric (in each symbol optimizing over $\psi$). This implementation is generally not realistic. A practical criterion is hard decoding of the M-QAM symbols of OFDM symbols, where the criterion function separates into a term for each symbol individually. In this case, the criterion function is the Euclidean distance between estimates  of the received signal point, $\widehat{S}_m$, at the output of the equalizer and the nearest point in the M-QAM constellation. Here, it is realistic to sequence through a quite large size $K$. Selection of the codebooks is a  design problem still to be considered. 

For the case where the Wiener PHN model is used, a systematic codebook design is presented in the next subsection. For other PHN models different codebook designs may perform better. However, the design for Wiener PHN may work well for other PHN models as well. This is illustrated in Section \ref{pll} where results based on a PHN modelled as Ornstein-Uhlenbeck process is presented.

\section{Codebook design for Weiner PHN \label{sec:3}}
The model introduced in \cite{Demir} and commonly employed in literature, where the PHN process is described by a Wiener process, is used. Under the Weiner model, the PHN sample at the $m$-th OFDM symbol is given by 
\begin{equation}\label{eq7}
\theta_m(n)=\sum_{i=0}^{mN+n}\varepsilon(i) =\psi_m + \sum_{l=0}^{n}\varepsilon(l)
\end{equation}
for $n=0,\ 1, \dots ,N-1$, where $\varepsilon(l)$ is a Gaussian random variable with zero mean and variance $\sigma^2_\varepsilon =2\pi\beta T/N$ (in radians$^2$), and $\beta T$ denotes the rate at which the PHN variance grows in one OFDM symbol. 
\begin{figure}
\centering
\psfrag{x}{\scriptsize{Time sample in an OFDM symbol}}
\psfrag{y}{\scriptsize{Phase Error (radians)}} 
\psfrag{data1}{$\theta(n)$}
\psfrag{data2}{$\hat{\theta}(n)$}
\includegraphics[scale=0.33]{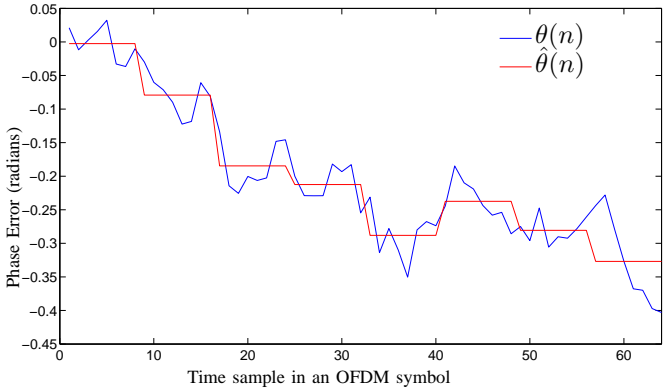}
\caption{A single Wiener PHN realization (blue) with $\sigma^2_\epsilon=2\pi(0.01/N)$ where $N=64$ and its approximation according to \eqref{eq8} for $J=8$.}
\label{fig2}
\end{figure}
\begin{figure}
\psfrag{h}{\tiny{$x_1$}}
\psfrag{w}{\tiny{$-x_1$}}
\psfrag{a}{\tiny{$x_2$}}
\psfrag{b}{\tiny{$x_r$}}
\psfrag{t}{\tiny{$-x_{2}$}}
\psfrag{u}{\tiny{$-x_{r}$}}
\psfrag{Q}{\tiny{$R_i$}}
\psfrag{R}{\tiny{$R_1$}}
\psfrag{J}{\tiny{$R_{i-1}$}}
\psfrag{K}{\tiny{$R_{\tiny{i+1}}$}}
\psfrag{D}{\tiny{$R_{\scriptsize{Q}}$}}
\psfrag{c}{$\cdots$}
\psfrag{d}{$\cdots$}
\includegraphics[scale=0.33]{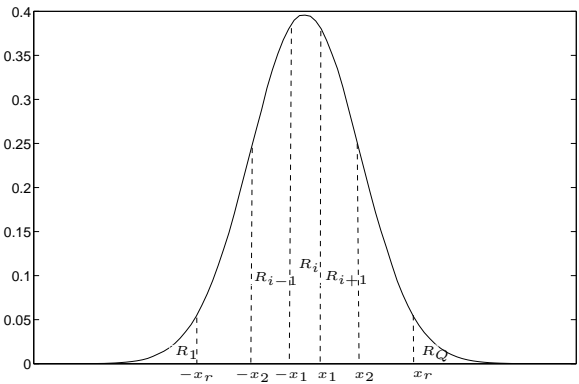}
\caption{Dividing the Gaussian sample space into $Q$ regions of equal areas, where $Q$ is an odd integer.}
\label{fig3}
\end{figure}
The objective is to construct a codebook containing a finite set of vectors which approximate the trajectories of possible Wiener PHN realizations within one OFDM symbol. The subscript $m$ in $\theta_m(n)$ and $\psi_m(n)$ is dropped for notational simplicity. Let the $N$ samples of the OFDM symbol be divided into  $J$ segments with $L=N/J$ samples per segment. Then $\hat{\theta}(n)$ is defined as a process which is constant in each segment. The value of the constants, $\lambda_j$, in each segment are set as the sample average of $\theta(n)$ in the corresponding segment. The PHN process is therefore approximated by $\hat{\theta}(n)$ given by
\begin{equation}\label{eq8}
\begin{split}
\hat{\theta}(n+jL)=\lambda_j&=\frac{1}{L}\sum_{\ell=0}^{L-1}\theta(\ell+jL)\\
&= \frac{1}{L}\sum_{\ell=0}^{L-1} \sum_{i=0}^{jL+\ell}\epsilon(i)  + \psi
\end{split}
\end{equation}
for $n=0, \cdots, L-1$ and $j=0, \cdots, J-1$. This is illustrated in Figure \ref{fig2} for $\sigma_\epsilon^2=2\pi(0.01/N)$, $N=64$ and $J=8$, where the samples within each segment are approximated by the sample mean of the elements in the segment. The process, $\hat{\theta}(n)$, is similar to a Weiner process where  each step, $X_{j}=\lambda_{j+1}-\lambda_{j}$, occurring every $L$ samples, is given by 
\begin{equation}\label{eq9}
\begin{split}
X_{j}&=\frac{1}{L}\sum_{\ell=0}^{L-1}\bigg(\theta((j+1)L+\ell) - \theta(jL+\ell)\bigg)\\
&=\frac{1}{L}\sum_{\ell=0}^{L-1} \sum_{i=\ell+jL+1}^{\ell+(j+1)L}\epsilon(i)
\end{split}
\end{equation}
where the second equality follows from \eqref{eq8}. The increments, $X_j$, are identically distributed Gaussian random variables with zero mean and variance $\sigma_x^2$ given by 
\begin{equation}\label{eq17k}
\begin{split}
\sigma_x^2 = E[X^2] &= \frac{1}{L^2}\sum_{\ell=0}^{L-1} \sum_{i=\ell+jL+1}^{\ell+(j+1)L}\sum_{v=0}^{L-1} \sum_{k=v+jL+1}^{v+(j+1)L}E[\epsilon(i)\epsilon(k)]\\
&=\frac{(2L^2+1)}{3L}\sigma_\epsilon^2
\end{split}
\end{equation}
where the auto-correlation $E[\epsilon(\ell)\epsilon(k)] = \sigma_\epsilon^2\delta(k-\ell)$ as well as arithmetic and geometric series are used to obtain the second equality. Note that for $L=1$, that is when the number of segments is equal to the number of samples, $J=N$, then $\sigma_x^2=\sigma_\epsilon^2$. The approximated Weiner PHN process can be re-written in terms of $X_{j}$ as  
\begin{equation}\label{eq10}
\hat{\theta}(jL+n)= \lambda_{j-1} + X_{j} +\psi = \sum_{\ell=0}^{j}X_\ell +\psi
\end{equation}
where $\lambda_{-1}=0$ and $X_0=0$ for $j=0$. 

Since the Gaussian sampling space is unbounded, the random variable $X$  is instead represented by a set of quantized samples, $\{\hat{x}_i\}_{i=1}^Q$, which represent $Q$ regions, $\{\mathcal{R}_i\}_{i=1}^Q$, in the Gaussian pdf with zero mean and variance $\sigma_x^2$. This implies that the increments at the consecutive segments, $j=1, \cdots, J-1$, are defined by a set $\{\hat{x}_i\}_{i=1}^Q$ which define $Q$ possible trajectories. That is, the second segment, $j=2$, has $Q$ possible increments which set $Q$ possible trajectories. In the next segment, there will be $Q$ possible increments on each trajectory from the previous segment such that there are $Q^{j-1}$ trajectories at the $j$-th segment. Therefore, at the $J$-th segment, there will be $K=Q^{J-1}$ trajectories representing quantized paths of the random walk process each of which are set as a vector entry in the codebook, ${\mathbf{\phi}}_{k=1}^K$. The number of code vectors, $K$, is therefore determined by the predefined set of quantization regions, $Q$, and the number of segments the symbol is divided into, i.e. $J$. 
\subsection{Defining Quantization Regions}
The quantization regions can be defined as a set of regions, $\{\mathcal{R}_i\}_{i=1}^Q$, which divides the Gaussian sample space  into $Q$ regions. For ease of development, the sample space is divided into a set of regions with equal areas (i.e. equiprobable). The regions are defined as  
\begin{equation}\label{eq11}
P\bigg[X\in \mathcal{R}_1\bigg]=\cdots= P\bigg[ X\in \mathcal{R}_Q \bigg]=\frac{1}{Q}
\end{equation}
\begin{figure}
\centering
\psfrag{x}{\scriptsize{Time sample in an OFDM symbol}}
\psfrag{y}{\scriptsize{Phase Error (radians)} }
\psfrag{data1}{{PHN process}}
\psfrag{data2}{\small{Set of trajectories}}
\includegraphics[scale=0.33]{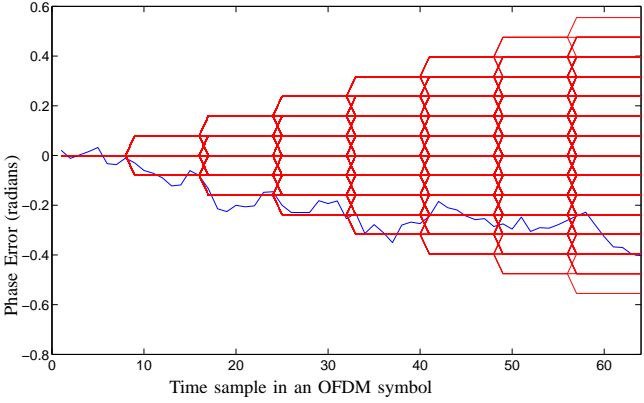}
\caption{A single Wiener PHN realization (blue) with $\sigma^2_\epsilon=2\pi(0.01/N)$ where $N=64$ and a finite set of $K=2187$ trajectories,* for $J=8$ and $Q=3$, which corresponds to possible trajectories that the PHN realization might take for equiprobable quantization. }
\label{fig4}
\end{figure}
where $P[X\in \mathcal{R}_i]$ implies the probability that the random variable $X$ is in the region $\mathcal{R}_i$. For the case when $Q$ is odd, the sample space is partitioned by $Q+1$ data points denoted by $\{-\infty, x_{-r}, \ldots,x_{-1}, x_{1}, \ldots, x_{r}, \infty \}$, as shown in Figure \ref{fig3}, where $x_{-\ell}=-x_{\ell}$, $\ell=1, \ldots, r$ and $r=(Q-1)/2$. The regions at the tails of the Gaussian pdf are given by $\mathcal{R}_{1}=[-\infty, x_{-r}[$ and $\mathcal{R}_{Q}=[x_{r}, \infty[$, while the regions in the middle are bounded by $[x_{\ell},x_{\ell+1}[$. The bounding points, $x_{\ell}$ are given by
\begin{equation}\label{eq12}
x_\ell = \sqrt{2\sigma_x^2} \erf^{-1}\bigg(\frac{2\ell-1}{Q}\bigg), \quad \ell=1, \cdots, \frac{Q-1}{2}               
\end{equation}
where $\erf^{-1}(\cdot)$ is the inverse error function. On the other hand, when $Q$ is even, the sample space is partitioned by $2r + 1$ data points denoted by $\{-\infty, x_{-r}, \cdots,x_{-1}, x_{0}, x_{1}, \cdots, x_{r}, \infty\}$ where $x_{0}=0$ and $x_{-\ell}=-x_{\ell}$ for $\ell=0, \cdots, r$ where $r=Q/2$. These points are given by 
\begin{equation}\label{eq13}
x_\ell = \sqrt{2\sigma_x^2} \erf^{-1}\bigg(\frac{2\ell}{Q}\bigg),  \quad \ell=0, \cdots, \frac{Q}{2}   .
\end{equation}
The plot in Figure \ref{fig4} is shown as an example to illustrate the process for a set of equiprobable quantization regions where $Q=3$ and $J=8$ while Figure \ref{fig4a} shows the trajectory that is closest to the PHN realization. The regions can also be defined as a set which divides the Gaussian sample space uniformly. This implies that  the codebook trajectory would have unequal probability. The development of this case is left for future work. 
\begin{figure}
\centering
\psfrag{x}{\scriptsize{Time sample in an OFDM symbol}}
\psfrag{y}{\scriptsize{Phase Error (radians)} }
\psfrag{l}{\tiny{PHN process}}
\psfrag{n}{\tiny{Selected trajectory } }
\psfrag{m}{\tiny{Selected trajectory (codebook with uniform quantization) } }
\psfrag{data1}{\tiny{PHN process}}
\psfrag{data2}{\tiny{Selected trajectory}}
\includegraphics[scale=0.33]{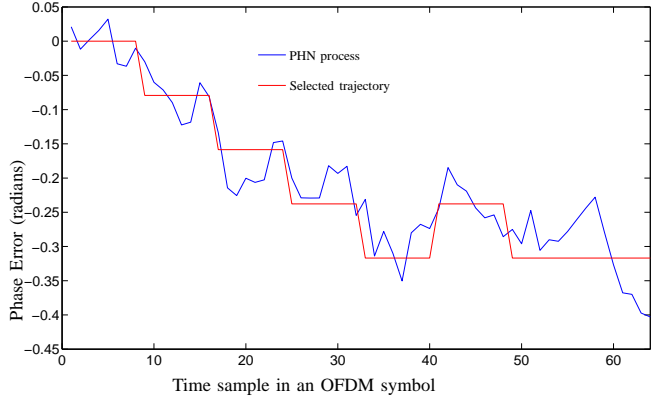}
\caption{The trajectory that best matches the PHN realization.}
\label{fig4a}
\end{figure}
\subsection{Quantization Points}
Given the quantization regions, $\mathcal{R}=\{R_1, \ldots, R_{Q}\}$, the quantization points, $\{\hat{x}_1, \ldots, \hat{x}_Q \}$, which represents each region is defined as the mean point within the region. That is,
\begin{equation}\label{eq14}
\begin{split}
\hat{x}_i =E\bigg[X|X\in R_i\bigg]&=\frac{\int_{R_i}xf_X(x)dx}{\int_{R_i}f_X(x)dx}\\
\end{split}
\end{equation}
for $i=1, \ldots, Q$, where $E[\cdot]$ is the expectation operator and $f(x)$ is the Gaussian distribution function. It can be shown that \eqref{eq14} provides the set which results in a minimum squared error quantization of the random variable $X$ with a pdf $f(x)$, \citep{Gersho}. Moreover, the expected value of quantization error $q_{\tiny{error}}=x-\hat{x}$ is zero, i.e. $E[\hat{x}]=E[x]$. For the Gaussian pdf with zeros mean and variance $\sigma_x^2=(2L^2+1)/3L\sigma_\epsilon^2$, \eqref{eq14} is given by
\begin{equation}\label{eq15}
\hat{x}_i = \frac{-\sigma_x}{0.5\sqrt{2\pi}}\bigg(\frac{\exp(-0.5x_{i+1}^2/\sigma_x^2)-\exp(-0.5x_{i}^2/\sigma_x^2)}{\erf(x_{i+1}/\sqrt{2}\sigma_x)-\erf(x_{i}/\sqrt{2}\sigma_x)}\bigg).
\end{equation}
It can be seen that the codebook will contain an all zeros entry when $Q$ is odd. This is attractive, since it means no correction will be done if the receiver would have no phase noise and the SNR is high. The plot in Figure \ref{fig4} employs \eqref{eq12} for $Q=3$ to define the quantization regions and \eqref{eq15} to  obtain the quantized point for each region. 
\begin{figure*}[t]
\centering
\captionsetup{justification=centering,margin=2cm}
\includegraphics[scale=0.63]{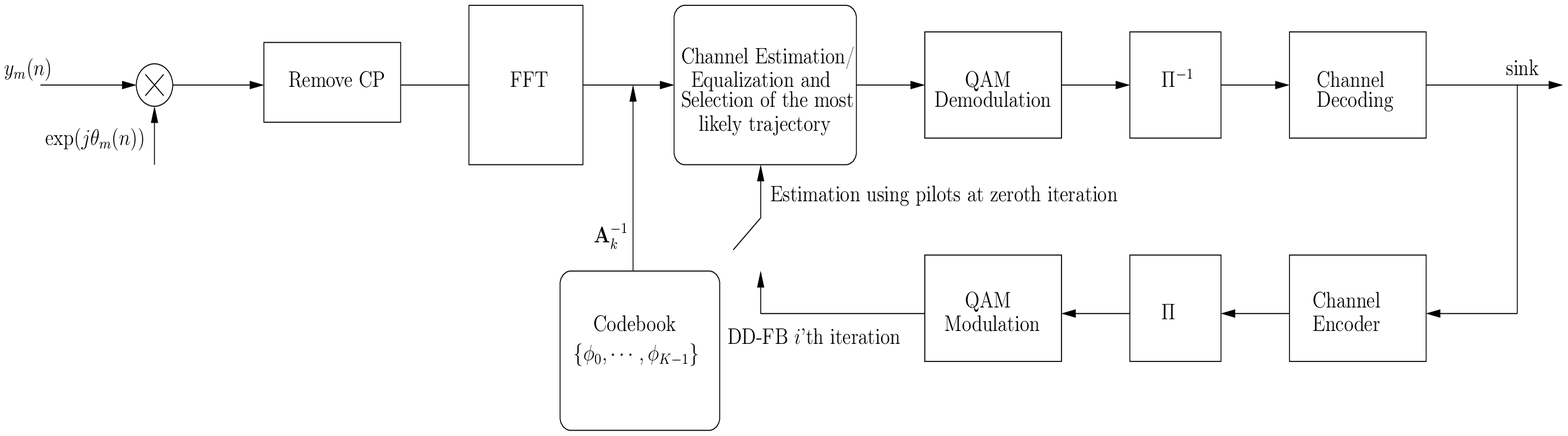}
\caption{Block diagram of the coded OFDM receiver with PHN compensation and channel estimation.}
\label{fig5}
\end{figure*}

\subsection{\mbox{MSE} Analysis \label{MSE}}
{In this section, the $\mbox{MSE}$ for the designed codebook is analysed. The objective is to provide a good indication of the performance of the designed codebook for a given $J$ and $Q$.} The $\mbox{MSE}$ between a random PHN realization and the approximate model introduced is given by
\begin{equation}\label{eq15_a}
\mbox{MSE}=E_{\theta(n)}[ \min_{k,\psi} \sum_n |\theta(n) - \psi- \phi_k(n)|^2 ]
\end{equation}
which is difficult to evaluate analytically. Therefore the MSE is evaluated by simulation. However, an approximate expression for the MSE, derived in Appendix \ref{sec:app}, is given by    
\begin{equation}\label{eq17}
\mbox{MSE} = \frac{(N+J)(N-J)}{6J}\sigma_\epsilon^2 + L(J-1)\sigma_q^2
\end{equation}
with $\sigma_q$ given by \eqref{eq17n}.  

Table \ref{tab1} shows the $\mbox{MSE}$ values obtained by Monte-Carlo simulations (denoted  $\mbox{MSE}_s$) normalized by the $\MSE$ for the CPE correction only (i.e. $\MSE$ for $J=1$). The number of PHN realizations employed were 5000 for every codebook with various combination of $J$ and $Q$ resulting in $K=Q^{J-1}$. The approximated theoretical $\mbox{MSE}_{\mbox{a}}$ values computed according to \eqref{eq17} are also given. It can be seen that the analytical and simulation results are similar indicating that the reasoning used in the analytical derivation in Appendix \ref{sec:app} is correct. The $\mbox{MSE}$ analysis assumes that the optimum code-book entry is chosen. In order to account also for the effect of noise and decision feedback errors, simulations of the full receiver for various codebook sizes are performed in Sec. \ref{sec:5}. These results confirm that the performance improves with increasing codebook size. The choice of codebook size will ultimately depend on the computational cost and performance requirements for the application at hand. However, it is noted that complexity grows rapidly with $Q$ and $J$, thus moderate numbers such as e.g. $Q=2$, $J=5$ or $Q=3$ and $J=4$ seems to provide a good compromise.
\begin{table*}[t]
\centering
\caption{Approximated $\mbox{MSE}$ for $N=64$ as is given by \eqref{eq17} and simulated $\mbox{MSE}$ denoted by $\mbox{MSE}_s$ for various values of $Q$ and $J$. The values are normalized by the $\MSE$ value for CPE correction only given by $\MSE=(N-1)(N+1)\sigma_\epsilon^2/6$.  \label{tab1}}
\begin{tabular}{|l|lll|lll|lll|lll|lll|} 
\toprule 
   & \multicolumn{3}{|c|}{$Q=2$} & \multicolumn{3}{|c|}{$Q=3$} & \multicolumn{3}{|c|}{$Q=4$} & \multicolumn{3}{|c|}{$Q=5$} & \multicolumn{3}{|c|}{$Q=6$} \\
    & K & $\mbox{MSE}_s$ & $\mbox{MSE}_{\mbox{a}}$
    & K & $\mbox{MSE}_s$ & $\mbox{MSE}_{\mbox{a}}$
    & K & $\mbox{MSE}_s$ & $\mbox{MSE}_{\mbox{a}}$
    & K & $\mbox{MSE}_s$ & $\mbox{MSE}_{\mbox{a}}$
    & K & $\mbox{MSE}_s$ & $\mbox{MSE}_{\mbox{a}}$   \\    \midrule

$J=1$  & 1 & 0.9967 & 1 & 1 & 0.9997 & 1& 1 & 0.9998 & 1 & 1 &  1.0021 & 1 & 1 & 1.0043 & 1  \\
$J=2$  & 1 & 0.7353  & 0.7387 & 2 &  0.6095  & 0.6247 & 3 & 0.5674 & 0.5765 & 4  &  0.5432 & 0.5519 &  5  & 0.5302 & 0.5376  \\
$J=4$  & 8  & 0.4479 & 0.4309 & 27  &  0.3488 & 0.3446 & 64 &   0.3076 & 0.3082 & 125 &   0.2878 & 0.2895 & 216  &  0.2770  & 0.2785 \\
$J=5$  & 16   &  0.4839 & 0.3607 & 81 & 0.2931  & 0.2855 & 256 & 0.2553 & 0.2538   & 625 &  0.2370  & 0.2375 & 1296 & 0.2274 & 0.2279\\
$J=8$  & 128   & 0.2631 & 0.2330 & 2187 & 0.1905 & 0.1813 & 16384 & 0.1601 & 0.1595   & 78125 &  0.1487  & 0.1482 & 279936 & 0.1441 & 0.1416\\ \hline
\end{tabular}
\end{table*}

\section{Implementation on OFDM System \label{sec:4}}
\subsection{Known Channel Response}
The received signal vector after removing cyclic prefix, $\mathbf{y}_m=[y_m(0), \cdots, y_m(N-1)]^T$, is multiplied by each of the $K$ trajectories $e^{-j\mathbf{\phi}_k}$ such that a new set of $K$ de-rotated OFDM signals is obtained at the receiver 
\begin{equation}\label{eq18}
\mathbf{\tilde{y}}_m^k=[e^{-j\phi_k(0)}y_m(0), \cdots, e^{-j\phi_k(N-1)}y_m(N-1)]^T.
\end{equation}  
An FFT operation on each de-rotated signal vector provides 
\begin{equation}\label{eq19}
\begin{split}
\mathbf{\widetilde{Y}}_m^k&=\mathbf{\widehat{A}}_{k}^{-1}\mathbf{Y}_m \\
&=\mathbf{\widehat{A}}_{k}^{-1}(\mathbf{A}_m\mathbf{S}^d_m\mathbf{H}_m + \mathbf{{W}}_m),
\end{split}
\end{equation}  
where the $N \times N$ circulant matrix $\mathbf{\widehat{A}}_{k}^{-1}$ is a matrix of DFT coefficients of $e^{-j\phi_k(n)}$. Frequency domain channel equalization is applied on $\mathbf{\tilde{Y}}_m^k$, and the criterion by which the trajectory that best approximates the PHN realization is given by 
\begin{equation}\label{eq21}
k^*= \arg\min_{k} \sum_{i=0}^{P-1}\bigg| \widehat{{S}}_m^k(\ell_i) - S_m(\ell_i)\bigg|^2. 
 \end{equation} 
where  $\ell_i \in (\ell_0, \dots , \ell_{P-1})$ denotes pilot subcarriers,  $\widehat{S}_m^k(\ell_i)$ is an element of the estimated symbol vector for the $k$-th trajectory given by 
\begin{equation}\label{eq21a}
\widehat{\mathbf{S}}_m^k=\eta_k^{-1}\mbox{diag}(\mathbf{H}_m)^{-1}\mathbf{\widetilde{Y}}_m^k
 \end{equation} 
with complex term $\eta_k$ that corrects for the effective CPE which includes the offset $\psi_m$ at each symbol, the DC level of the codebook vector, $\phi^k$, and the accumulated quantization error. It is given by  
 \begin{equation}\label{eq22}
\eta_k= \frac{\sum_{i=0}^{P-1} S_m^*(\ell_i)\widetilde{Y}_m^k(\ell_i)/H_m^k(\ell_i)}{ \sum_{i=0}^{P-1} |S_m(\ell_i)|^2}.
 \end{equation} 
In a DD-FB loop in which the output of the channel decoder can be exploited, non-pilot symbols can also be used such that the criterion in \eqref{eq22} takes into account the decoded symbols as well
\begin{equation}\label{eq23}
k^*= \arg\min_{k} \sum_{l=0}^{N-1}\bigg| \widehat{{S}}_m^k(l) - S_m(l)\bigg|^2. 
 \end{equation} 
for  $\eta_k$ given by    
\begin{equation}\label{eq24}
\eta_k= \frac{\sum_{l=0}^{N-1} S_m^*(l)\widetilde{Y}_m^k(l)/H_m^k(l)}{ \sum_{i=0}^{N-1} |S_m(l)|^2}.
 \end{equation} 
The symbol sent to the decoder is then given by the index corresponding to $\widehat{\mathbf{S}}_m^{k^*}.$

\subsection{Combined Channel Estimation and ICI Cancellation }
Since the channel is not always known at the receiver, it needs to be estimated. 
Assuming the channel remains stationary within the time period of an OFDM symbol, for a codebook of size  $K$, there will be $K$ received signal candidates, $\mathbf{\tilde{Y}}^k_m$,  which are given by \eqref{eq19} where the channel response, $H_m(k)$, is considered to be unknown. The MMSE estimate of the channel frequency response vector, $\widehat{\mathbf{H}}_m^k$, employing the $k$-th trajectory of the codebook is then given by \cite{Beek} 
\begin{equation}\label{eq25}
\widehat{\mathbf{H}}_m^k=E[\mathbf{H}_m(\mathbf{H}_m)^H]\bigg(E[\mathbf{H}_m(\mathbf{H}_m)^H]+\frac{\sigma^2_w}{E_s}\mathbf{I}_{N}\bigg)^{-1}\widehat{\mathbf{H}}^k_{\mbox{\tiny{LS}},m}
\end{equation}
where $E[\mathbf{H}_m (\mathbf{H}_m)^H]$ is the autocorrelation matrix of the channel frequency response for the given statistical model of the channel and $\widehat{\mathbf{H}}_{\mbox{\tiny{LS}},m}^k$ is the least squares estimate of $\mathbf{H}^k_m$ given by 
\begin{equation}\label{eq26}
\widehat{\mathbf{H}}^k_{\mbox{\tiny{LS}},m} = \mathbf{S}_m^{-1}\widetilde{\mathbf{Y}}_m^k = \bigg[\frac{\widetilde{Y}^k_m(0)}{S_m(0)}, \cdots, \frac{\widetilde{Y}^k_m(N-1)}{S_m(N-1)}\bigg]^T.
\end{equation}
However, since not all transmitted symbols are known at the receiver, $N_p$ pilot symbols, fitted evenly among the $N$ subcarriers, are employed to obtain a least squares estimate of the channel frequency response at the pilot positions. For the sake of convenience, the LS estimate at pilot positions are denoted by the vector, $\widehat{\mathbf{H}}_{\mbox{\tiny{LS}},m}^{k,p}$. Therefore, the MMSE estimate of the channel frequency response, for $k$-th trajectory in the codebook, is initially obtained based on pilot symbols     
\begin{equation}\label{eq27}
\widehat{\mathbf{H}}_m^k=E[\mathbf{H}_m(\mathbf{H}_m^p)^H]\bigg(E[\mathbf{H}_m^{p}(\mathbf{H}_m^{p})^H]+\frac{\sigma^2_w}{E_s}\mathbf{I}_{N_p}\bigg)^{-1}\widehat{\mathbf{H}}_{\mbox{\tiny{LS}},m}^{k,p}
\end{equation}
where $(\cdot)^p$ denotes a vector whose elements are positioned at pilot subcarriers. A set of $K$ channel frequency response estimates, $\{ \widehat{\mathbf{H}}_m^k \}_{k=1}^K$, corresponding to each trajectory in the codebook is available at the receiver.  Equation \eqref{eq23} is used to determine the trajectory which closely matches the PHN realization, and the corresponding channel estimate is then used to demodulate and obtain rough decisions on the data symbols. After channel decoding, a decision feedback technique is then employed to obtain \eqref{eq26} based on rough decisions on the symbols containing pilot and decision feedback symbols, $\overline{\mathbf{S}}_m$, which is then used in \eqref{eq25} to compute an MMSE estimate of $\mathbf{H}^k_m$ with improved accuracy.  

For deeply faded channels, estimation accuracy during symbol $m$ can be improved by including previously decoded symbols, i.e., $\mathbf{\overline{S}}_{m-1}, \mathbf{\overline{S}}_{m-2}, \cdots, \mathbf{\overline{S}}_{m-D+1}$, in the estimation vector. 
Therefore, taking into account $D$ previously decoded OFDM symbols, the MMSE estimator in \eqref{eq27} becomes  
\begin{equation}\label{eq28}
\widehat{\mathbf{H}}_m^k=E[\mathbf{H}_m(\mathbf{H})^H]\bigg(E[\mathbf{H}(\mathbf{H})^H]+\frac{\sigma^2_w}{E_s}\mathbf{I}_{DN}\bigg)^{-1}\widehat{\mathbf{H}}_{\mbox{\tiny{LS}}}
\end{equation}
where $\mathbf{H}=[\mathbf{H}_m, \mathbf{H}_{m-1}, \cdots, \mathbf{H}_{m-D+1}]^T$ and $\widehat{\mathbf{H}}_{\mbox{\tiny{LS}}}=[\widehat{\mathbf{H}}_{\mbox{\tiny{LS}},m}^{
k_{m}}, \widehat{\mathbf{H}}_{\mbox{\tiny{LS}},m-1}^{k_{m-1}}, \cdots, \widehat{\mathbf{H}}_{\mbox{\tiny{LS}},m-D+1}^{k_{m-D+1}}]^T$. On the initial run of the $m$-th OFDM symbol, symbols at pilot subcarriers are used in order to compute \eqref{eq27}, after which the criterion in \eqref{eq21} is used to choose the best match index, $k^*$, for the trajectory which closely matches the PHN realization, $\mathbf{\phi}^{k^*}$, as well as the corresponding channel estimate, $\widehat{\mathbf{H}}_m^{k^*}$. Pilot and decision feedback symbols can then be used in the next iteration to obtain a better estimate, $\widehat{\mathbf{H}}_m^k$ using \eqref{eq25} or \eqref{eq28} if using $L$ previous OFDM symbols and \eqref{eq23} to obtain $k^*$. Joint channel equalization and PHN compensation is employed on the received signal $\mathbf{Y}_m$ such that the symbols sent to the decoder are given by
\begin{algorithm}
\centering
\caption{The proposed receiver algorithm with no DD-FB for uncoded system and known channel}\label{alg1}
\begin{algorithmic}[1]
\Procedure{ }{}
\State Received $m$-th OFDM symbol
   \For{$k=0:K-1$}
        \State $\mathbf{\widetilde{Y}}_m^k=\mathbf{\widehat{A}}_{k}^{-1}\mathbf{Y}_m$        
      	\State Compute $\eta_k$ in \eqref{eq22} using pilot symbols
      	 \State Compute $F(k)=\sum_{i=0}^{P-1}|\widehat{S}_m^k(\ell_i)   - S_m(\ell_i)|^2$    
   \EndFor
      	\State Compute $k^*= \arg\min_{k} F(k)$ in  \eqref{eq21} to obtain $k^*$
          \State \textbf{return} $\widehat{\mathbf{S}}_m^{k^*}$
\EndProcedure
\end{algorithmic}
\end{algorithm}
\begin{equation}\label{eq29}
\widehat{\mathbf{S}}_m^{k^*} = \eta_{k^*}^{-1}\mbox{diag}(\widehat{\mathbf{H}}_m^{k^*})^{-1}\widetilde{\mathbf{Y}}_m^{k^*}
\end{equation} 
where $k^*$ denotes the index of the most likely trajectory in the codebook obtained from \eqref{eq23}. Figure \ref{fig5} shows the OFDM receiver with the proposed PHN compensation and channel estimation scheme. An outline of the proposed algorithm for a known channel response without DD-FB loop is presented in Algorithm \ref{alg1}. In Algorithm \ref{alg2}, the algorithm for the combined channel frequency response estimation and PHN compensation scheme is presented for a coded frame of $M$ OFDM symbols.

\begin{algorithm}
\centering
\caption{The proposed algorithm for channel estimation and PHN compensation with DD-FB and channel coding}\label{alg2}
\begin{algorithmic}[1]
\Procedure{ }{}
\State Received frame
\Do { \em{Iteration on DD-FB loop}  }
 \Do { \em{for each OFDM symbol in the frame} }   
    \For{$k=0:K-1$}
        \State $\mathbf{\widetilde{Y}}_m^k=\mathbf{\widehat{A}}_{k}^{-1}\mathbf{Y}_m$
      \If{Zeroth iteration}
        \State Channel estimation using pilots, \eqref{eq27}  
      	\State Compute $\eta_k$ in \eqref{eq22} using pilot symbols  
      	\State Detected symbol for each $k$ using \eqref{eq29}     	
        \State Compute $F(k)=\sum_{i=0}^{P-1}| \widehat{S}_m^k(\ell_i)-$  \NoNumber{ $S_m(\ell_i)|^2$}
       \Else
        \State Channel estimation employing \eqref{eq25}  
      	\State Compute $\eta_k$ using \eqref{eq24}     	
        \State Compute $F(k)=\sum_{i=0}^{N-1}|\widehat{S}_m^k(\ell_i) -$  \NoNumber{  $S_m(\ell_i)|^2$}
       \EndIf
   \EndFor
        \State Compute $k^*= \arg\min_{k} F(k)$ in \eqref{eq21} (zeroth iter-   \NoNumber{ ation) or \eqref{eq23} to obtain $k^*$}
        \State Select $\mathbf{\widehat{S}}_m^{k^*}$ 
    \doWhile{ \textit{End of frame} }
    
    \State Decode frame and deinterleave
    \State Interleave, channel coding and modulation frame
    \State DD-FB frame $\mathbf{\overline{S}}=[\mathbf{\overline{S}}_{0}, \cdots, \mathbf{\overline{S}}_{m}, \cdots, \mathbf{\overline{S}}_{M-1}]$
   \doWhile{ \textit{End of iteration} }    
\EndProcedure
\end{algorithmic}
\end{algorithm}
\section{Simulation Results \label{sec:5}}
In this section, simulation results are given to show the BER performance of the proposed technique under Wiener PHN. Two sets of simulation results are presented where in the first set, the performance of the algorithm in Algorithm \ref{alg1} for an uncoded system is presented under AWGN channel. In the second set, performance of the algorithm in Algorithm \ref{alg2} is shown for a Rayleigh fading channel. Simulation results are presented for 16-QAM and 64-QAM modulation schemes with $N=64$, $N_{cp}=16$ and $N_p=8$ pilot subcarriers. The codebook size for the results presented is $K=27$, in which $Q=3$ and $J=4$, which from the analysis in Sec. \ref{MSE} is a good compromise between complexity and performance. Perfect timing and frequency synchronization is assumed at the receiver.

\subsection{Uncoded System, AWGN Channel}
\begin{figure}
\centering
\psfrag{x}{\scriptsize{$E_b/N_0$(dB)} }
\psfrag{y}{\scriptsize{BER}}
\psfrag{data1}{\tiny{\cite{Munier}, 3 DD-FB iterations (64-QAM)} }
\psfrag{data2}{\tiny{Proposed method (64-QAM),  no DD-FB}}
\psfrag{data3}{\tiny{AWGN channel no PHN (64-QAM)} }
\psfrag{data4}{\tiny{\cite{Munier}, 3 DD-FB iterations (16-QAM)} }
\psfrag{data5}{\tiny{Proposed method (16-QAM)}, no DD-FB }
\psfrag{data6}{\tiny{AWGN channel no PHN (16-QAM)} }
\includegraphics[scale=0.38]{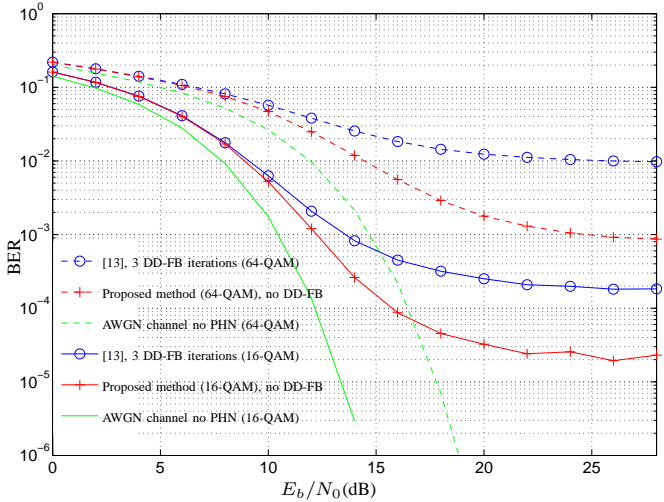}
\caption{BER performance of proposed algorithm compared to \cite{Munier} for 16-QAM and 64-QAM uncoded OFDM under AWGN channel and PHN. Corresponding performance plots for an ideal case (i.e. AWGN channel no PHN) is also given. $BT=0.01$. } 
\label{fig6}
\end{figure}

Considering the AWGN channel, the proposed Algorithm \ref{alg1} was compared with the PHN compensation method presented in \cite{Munier}, employing 3 iterations in the DD-FB. In the proposed method no DD-FB iterations are employed. The rate of growth of the PHN variance $\beta T$ is set to $0.01$, and 16-QAM OFDM and 64-QAM OFDM were considered without channel coding. The result is given in Figure \ref{fig6}. For the 64-QAM system (dashed line), the proposed method is $2.5$dB from the ideal PHN free case at a BER of $10^{-2}$, while the method in \cite{Munier} is $9$dB from the ideal case, i.e., an improvement of $6.5$dB. In the 16-QAM system (solid line), the proposed method is $2$dB from the ideal PHN free case at a BER of $10^{-3}$, while the method in \cite{Munier} is $3$dB from the ideal case.

\subsection{Coded System with Fading Channel\label{sec:5a}}
A 10 tap Rayleigh fading channel with parameters specified in Table.\ref{tab3} is used. Moreover, 
a 1/2 rate convolutional code with a constraint length of 7 is used on the input bit stream, which is then passed on to a bit interleaver over 20 OFDM symbols. A Viterbi decoder is used at the receiver. The PHN variance growth rate, $\beta T$, is set to $0.01$ for 16-QAM OFDM such that the RMS in degrees is $180\sqrt{2\pi\beta T}/\pi=14.4^0$; while for 64-QAM, $\beta T$ is set to $0.005$. Higher order modulation schemes such as 64-QAM are known to be affected particularly worse by PHN. The simulated channel, also employed in \cite{Munier}, is given for a normalized amplitude by \cite{Hoeher} 
\begin{table}[t]
\centering
\caption{summary of the Rayleigh channel parameters  \label{tab3}}
\begin{tabular}{ll}
   \\ \hline
Carrier Frequency, $f_c$    & 5 GHz    \\
Sampling rate, $f_s=\frac{1}{T_s}$         & 25 MHz  \\
rms delay spread, $\frac{\tau_{\scriptsize{rms}}}{T_s}$    & 120ns   \\
Speed, $v$                & 7 km/h \\
Channel taps, $L$        & 10 \\\hline
\hline
\end{tabular}
\end{table}
\begin{equation}\label{eq30}
H_m(k)=\sum_{i=0}^{L-1}\phi(\tau_i)e^{j(\varphi_i + 2\pi f_i^D m + 2\pi \frac{k\tau_i}{N})}
\end{equation}
where $\alpha_i$, $\varphi_i$, $f_i^D$ and $\tau_i$  are respectively the amplitude, phase,  Doppler frequency and time delay of the $i$-th propagation path. The time correlation for the given channel model is defined as 
\begin{equation}\label{eq31}
R_t(m-n)=E[H_m(k)H^*_n(k)].
\end{equation}
Assuming uniformly distributed angles of arrival at the mobile station, the correlation follows the well known Jakes model \cite{jakes} such that $R_t(m-n)=J_0(2\pi f_{max}^D(m-n))$, where $J_0(\cdot)$ denotes the zeroth order Bessel function of the first kind, and $f_{max}^D=(vf_c/c)T$ is the maximal Doppler frequency normalized by the OFDM symbol duration, $T$. Assuming an exponentially decaying power delay profile with normalized RMS delay spread $\tau_{\text{rms}}$, \citep{Sandell}, the correlation between different frequency bins is given by

\begin{equation}\label{eq32}
\begin{split}
R_f(k-l)&=E[H_m(k)H^*_m(l)]\\
&=\frac{1-e^{-L(1/\tau_{rms}+j2\pi(k-l))}}{(1-e^{-L/\tau_{rms}})(1+j2\pi(k-l)\tau_{rms})}
\end{split}
\end{equation}
which satisfies $E[|H_m(k)|^2]=1$.

\subsubsection{Channel fully known at the Receiver}
Assuming that the frequency response of the channel is known at the receiver, the performance of the proposed algorithm is presented and compared with previously proposed techniques in \cite{Petrovic} and \cite{Munier}. 

In \cite{Petrovic}, a Fourier series representation is used to approximate the PHN realization at the $m$-th OFDM symbol, thereby aiming to suppress all the influence of the PHN. CPE compensation and symbol estimates are initially obtained using pilot subcarriers, after which a DD-FB loop is employed for MMSE estimation of the vector of Fourier coefficients of the PHN realization which is subsequently used for ICI compensation. Similarly, in \cite{Munier}, a two stage cancellation technique is employed in which the CPE is corrected,  followed by ICI cancellation using decision feedback symbols. The ICI cancellation technique in \cite{Munier} is derived  based on power series expansion of the PHN over an OFDM symbol. 

The results in our approach were obtained using Algorithm \ref{alg2}, but where step 8 and 12 are omitted since the channel is already known. The results are shown by the solid curves in Figure \ref{fig7} and Figure \ref{fig8} for 16-QAM and 64-QAM respectively. 

First, in Figure \ref{fig7} (solid line), it can be seen that the performance of the proposed algorithm without employing  DD-FB is $0.5$dB from the ideal case (no PHN) for a BER of $10^{-4}$. On the other hand, for 64-QAM, Figure \ref{fig8} shows that the performance of the proposed algorithm is $2$dB from the ideal plot for a BER of $10^{-4}$. In both Figures, the proposed method shows performance gain in BER compared to the previously proposed methods in \cite{Petrovic}, which employs a single iteration on the DD-FB, and the method proposed in \cite{Munier},  which is shown for 2 and 3 DD-FB iterations respectively for 16-QAM and 64-QAM. It should be noted that no DD-FB loop is performed in the proposed approach.
\subsubsection{Unknown Channel Response}
For the case where the channel is unknown, joint channel estimation and PHN compensation is employed according to Algorithm \ref{alg2}. Performance of the techniques in \cite{Munier} and \cite{Corvaja} is also presented for comparison. 

In \cite{Corvaja}, FFT interpolation is employed using the least squares estimates of the joint channel frequency response and CPE at pilot positions to obtain estimates over all subcarriers. The method also proposes estimating the CPE by averaging the phase displacements on pilot subcarriers which is  used to remove its influence from the estimated channel frequency response. Previously decoded symbols are then employed for least squares (LS) estimation of the vector of Fourier coefficients of the PHN realization for ICI compensation using previous channel and symbol estimates. Three sets of results are presented for the method in  \cite{Corvaja}. The first two results consider estimation of only one and three Fourier coefficients of the PHN realization closest to the DC for ICI compensation thus requiring less computational complexity. The third result requires inverting a $64\times64$ Toeplitz matrix to estimate the entire vector of Fourier coefficients of the PHN realization for ICI compensation. 

\begin{figure}
\centering
\psfrag{x}{\scriptsize{$E_b/N_0$(dB)} }
\psfrag{y}{\scriptsize{BER}}
\psfrag{data12}{\tiny{Full channel knowledge and no PHN}}
\psfrag{data11}{\tiny{Proposed method, without DD-FB loop (known channel)} }
\psfrag{data10}{\tiny{\cite{Munier}, 2 DD-FB iterations (known channel)} }
\psfrag{data9}{\tiny{ICI-cancellation Method in \cite{Petrovic} (known channel) }}
\psfrag{data8}{\tiny{Proposed method, 2 DD-FB iterations with channel estimation} }
\psfrag{data7}{\tiny{Proposed method without DD-FB with channel estimation} }
\psfrag{data6}{\tiny{\cite{Munier}, 8 DD-FB iterations with channel estimation} }
\psfrag{data5}{\tiny{\cite{Munier}, 2 DD-FB iterations with channel estimation} }
\psfrag{data4}{\tiny{\cite{Munier}, without DD-FB with channel estimation} }
\psfrag{data3}{\tiny{\cite{Corvaja}, full matrix} }
\psfrag{data2}{\tiny{\cite{Corvaja}, partial matrix 3 } }
\psfrag{data1}{\tiny{\cite{Corvaja}, partial matrix 1 } }
\includegraphics[scale=0.37]{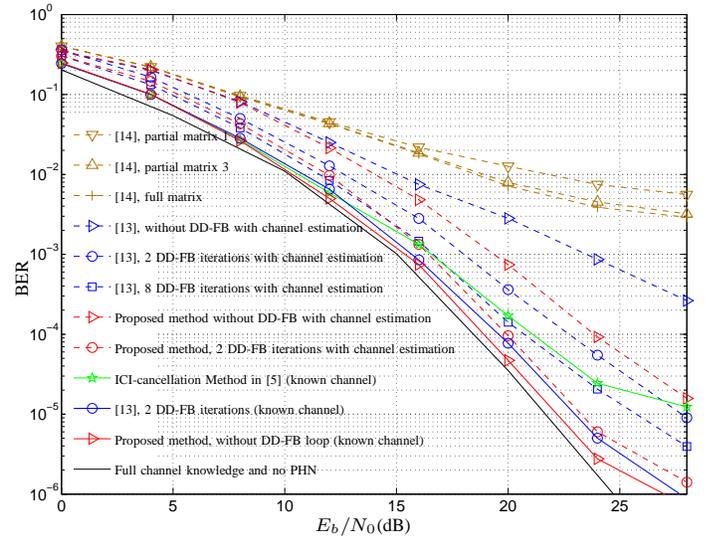}
\caption{BER performance of the proposed algorithm compared to \cite{Munier} and \cite{Corvaja} for 16-QAM OFDM. Performance plot for the ideal case (i.e. perfect channel knowledge and no PHN) is also given. ICI cancellation technique in \cite{Petrovic} is also shown for perfectly known channel. $BT=0.01$} 
\label{fig7}
\end{figure}

\begin{figure}
\centering
\psfrag{x}{\scriptsize{$E_b/N_0$(dB)} }
\psfrag{y}{\scriptsize{BER}}
\psfrag{data9}{\tiny{Full channel knowledge and no PHN }}
\psfrag{data8}{\tiny{Proposed method without DD-FB loop (known channel)} }
\psfrag{data7}{\tiny{\cite{Munier} 3 DD-FB iterations (known channel)} }
\psfrag{data6}{\tiny{ICI-cancellation Method in \cite{Petrovic} (known channel) }}
\psfrag{data5}{\tiny{Proposed method, 10 DD-FB iterations with channel estimation} }
\psfrag{data4}{\tiny{Proposed method, 3 DD-FB iterations with channel estimation} }
\psfrag{data3}{\tiny{\cite{Munier}, 10 DD-FB iterations with channel estimation} }
\psfrag{data2}{\tiny{\cite{Munier}, 3 DD-FB iterations with channel estimation} }
\psfrag{data1}{\tiny{\cite{Corvaja}, full matrix } }
\includegraphics[scale=0.38]{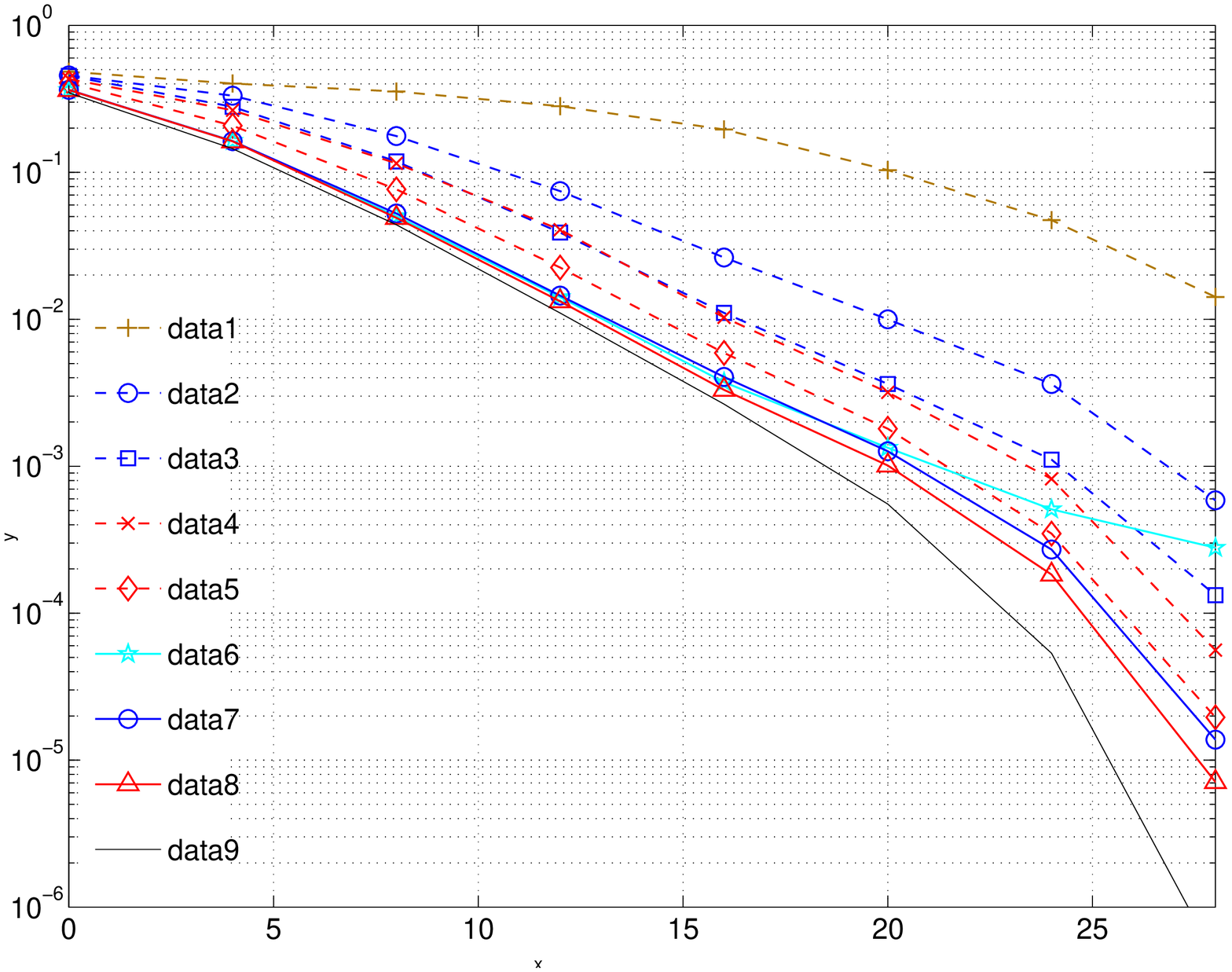}
\caption{BER performance of proposed algorithm compared to \cite{Munier}  and \cite{Corvaja} for 64-QAM OFDM. Performance plot for the ideal case (i.e. perfect channel knowledge and no PN) is also given. ICI-cancellation technique in \cite{Petrovic} is also shown for perfectly known channel. $BT=0.005$}
\label{fig8}
\end{figure}

The proposed method as well as the method in \cite{Munier} use $D=3$ past OFDM symbols as described in \eqref{eq23} to enhance performance of the channel estimation. Performance plots for the unknown channel case are shown with the dashed curve in Figure \ref{fig7} and Figure \ref{fig8} for 16-QAM and 64-QAM respectively. In both Figure \ref{fig7} and Figure \ref{fig8}, the performance of the proposed technique shows improved performance, in BER, over the previously proposed methods in \cite{Munier}, \cite{Corvaja}.

In Figure \ref{fig7}, the proposed method is $1.5$dB from the ideal case, (no PHN and known channel response) at a BER of $10^{-4}$ employing only 2 iterations on the DD-FB loop. On the other hand, the method in \citep{Munier}, shown for 8 iterations, is $2.5$dB from the ideal case while the method in \citep{Corvaja} is far from the ideal case. 

For the 64-QAM system, Figure \ref{fig8} shows that the proposed method provides improved performance compared to \citep{Munier} and the method in \citep{Corvaja}. With 10 iterations, the proposed methods is $3$dB from the ideal case at a BER of $10^{-3}$, while the method in \cite{Munier} is $5$dB from the ideal case for the same number of iterations. 
\begin{figure}
\centering
\psfrag{x}{\scriptsize{$E_b/N_0$(dB)} }
\psfrag{y}{\scriptsize{BER}}
\psfrag{data4}{\tiny{$K=27$, $Q=3$, $J=4$}}
\psfrag{data5}{\tiny{$K=625$, $Q=5$, $J=5$}}
\psfrag{data6}{\tiny{$K=2187$, $Q=3$, $J=8$}}
\psfrag{data1}{\tiny{$K=1$, $Q=1$, $J=1$}}
\psfrag{data2}{\tiny{$K=4$, $Q=5$, $J=2$}}
\psfrag{data3}{\tiny{$K=16$, $Q=2$, $J=5$}}
\psfrag{a}{\tiny{with channel estimation}}
\psfrag{b}{\tiny{known channel}}
\psfrag{d}{\tiny{$\beta T=0.01$ (red)}}
\psfrag{c}{\tiny{$\beta T =0.05$ (blue)}}
\includegraphics[scale=0.38]{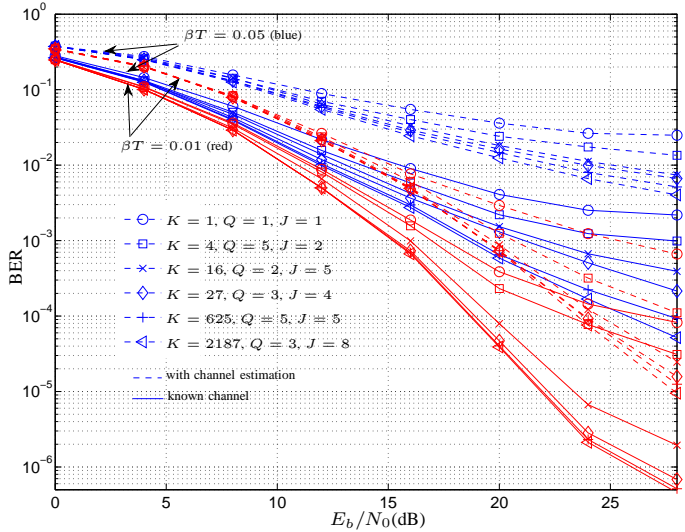}
\caption{BER performance of the proposed algorithm on 16-QAM coded OFDM for various number of codebook size.}
\label{fig_aa}
\end{figure}
\subsubsection{Performance with respect to size of $K$}
\indent In Figure \ref{fig_aa}, simulation results are presented showing the BER performance of the proposed algorithm for various codebook size $K$ and for $\beta T=0.01$ and $\beta T=0.05$. The presented result does not employ any DD-FB loop. A 16-QAM OFDM system with the same description as in Sec. \ref{sec:5a} is considered assuming known and unknown Rayleigh fading channel. It can be seen that the BER improves a $K$ grows. However, the relative improvement as $K$ grows beyond $K=27$ is very small. For $\beta T=0.01$, BER improves only slightly at high $\SNR$ in both cases where the channel is assumed known and where the channel is estimated as $K$ increases from 27. The relative improvement in BER as $K$ increases is more noticeable for $\beta T=0.05$, even though it is still small. 
\subsubsection{Sensitivity to input parameters}
\begin{figure}
\centering
\psfrag{x}{\scriptsize{$\widehat{f}_{max}^D$} }
\psfrag{y}{\scriptsize{BER}}
\psfrag{data1}{\tiny{$\widehat{\beta T}=0$}}
\psfrag{data2}{\tiny{$\widehat{\beta T}=0.01$}}
\psfrag{data3}{\tiny{$\widehat{\beta T}=0.03$}}
\psfrag{data4}{\tiny{$\widehat{\beta T}=0.05$}}
\psfrag{data5}{\tiny{$\widehat{\beta T}=0.07$}}
\psfrag{a}{\tiny{Proposed method}}
\psfrag{b}{\tiny{Method in \citep{Munier}}}
\psfrag{c}{\tiny{Method in \citep{Corvaja}}}
\psfrag{cc}{\tiny{ $f_{max}^D$}}
\includegraphics[scale=0.38]{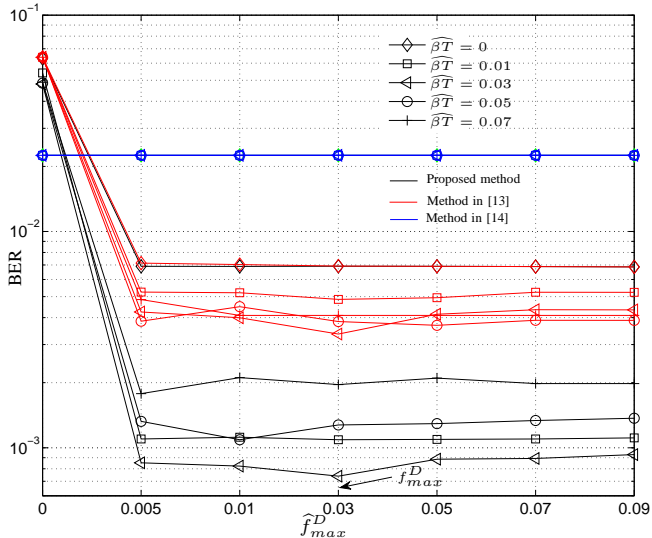}
\caption{BER performance of the proposed algorithm (black line) and the method in \citep{Munier} (red line) as a function of the input doppler frequency $\widehat{f}_{max}^D$ for various input values of $\beta T$  in a 16-QAM OFDM on a Rayleigh channel where the true $f_{max}^D=0.03$, $\beta T=0.03$ and $\SNR=20$dB.}
\label{fig_bb}
\end{figure}

\indent The plots in Figure \ref{fig_bb} show the sensitivity of the proposed algorithm and the algorithm in \citep{Munier} to the estimated channel and PHN parameters relative to the true parameters. The rate at which the PHN variance grows is set to $BT=0.03$ and  the normalized maximal Doppler frequency is set to $f_{max}^D=0.03$. All other Rayleigh channel  parameters are kept the same as in Table \ref{tab3}. Moreover, the size of the codebook was $K=27$ with $Q=3$ and $J=4$ and $\SNR=20$dB. Both the proposed method and the method in \cite{Munier} use the past two symbols together with the current symbol $D=3$ for channel estimation.  

It can be seen that the sensitivity of the proposed algorithm as well as the algorithm in \cite{Munier} to the input values of the normalized maximal Doppler frequency, $\widehat{f}_{max}^D$ is low except for $\widehat{f}_{max}^D=0$. However, for higher value of $D$, the BER performance is expected to be very sensitive to the input value of $\widehat{f}_{max}^D$. However, it can be seen that the performance of both the proposed algorithm and \cite{Munier}  is noticeably sensitive to the input values of $\widehat{\beta T}$. On the other hand, the method in \cite{Corvaja} does not depend at all on the channel and PHN statistics and thus the performance remains unaffected by the estimated input parameters. 
\subsubsection{PLL Oscillator \label{pll}}
The above results have been provided assuming a free running oscillator in which the PHN is modelled by the Weiner process and based on which the codebook is derived. In order to demonstrate the applicability for a PHN model other than the Weiner process, we employ the proposed method using the Ornstein-Steinbeck process \cite{Gardiner} to model the PHN. The Ornstein-Uhlenbeck process is used to model the PHN at the output of a voltage controlled oscillator (VCO) in a phase-locked loop (PLL) \cite{Petrovic}. 

\begin{figure}
\centering
\psfrag{x}{\scriptsize{Sample} }
\psfrag{y}{\scriptsize{Phase Error (radians)}}
\psfrag{data1}{\tiny{Free running Oscillator  }}
\psfrag{data2}{\tiny{Charge Pump PLL} }
\includegraphics[scale=0.38]{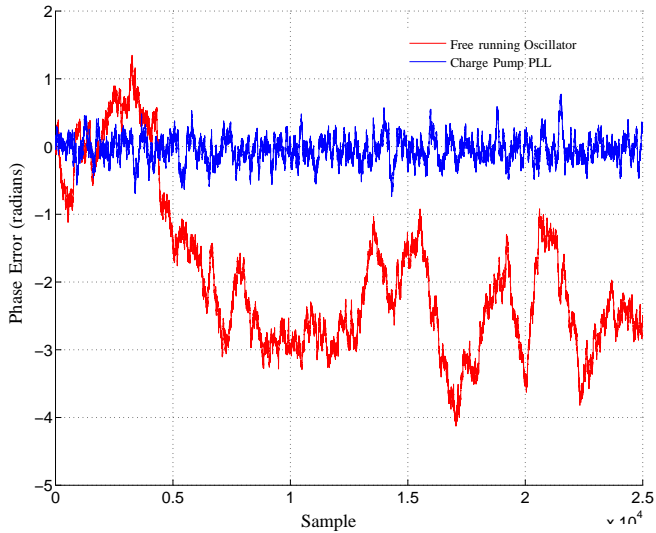}
\caption{PHN samples, $\theta(n)$, simulated as an out from a free running oscillator (Weiner process) and a charge Pump PLL (Ornstein-Uhlenbeck process).}
\label{fig_smpl}
\end{figure}

In Figure  \ref{fig_smpl}, a PHN realization, $\theta(n)$, for a free running oscillator  and a  PLL is shown. It can be seen that the PHN from PLL (Ornstein-Uhlenbeck process) is stable around the mean while the PHN from a free running oscillator grows over time. 

A detailed analysis of PLL circuit parameters and the associated PHN process generated is given in \citep{Mehrotra}. In Figure \ref{fig_smpl}, the PHN variance of the Weiner process (free running VCO) grows at a rate of $(\beta T)_{vco}=0.05$ while more parameters are required to model the Ornstein-Uhlenbeck process. A Charge pump PLL is considered which is set to have a 3dB bandwidth of $f_{lp}=20$kHz for the low pass filter, $f_{pd}=20$kHz for the phase detector and $f_{pll}=100$kHz for the PLL itself. The PHN variance of the VCO and the reference oscillator grow at a rate of $\beta T_{vco}=0.01$ and $(\beta T)_{ref}=2^{-9}$ respectively. Moreover, the center frequency of the VCO is $f_c=5$GHz while the center frequency of the reference oscillator is $f_{ref}=100$MHz and the sampling frequency is $f_s=25$MHz.  A 16-QAM coded OFDM with the same description as in Sec. \ref{sec:5a} is considered assuming unknown Rayleigh fading channel so that channel estimation is employed along PHN compensation. 

\begin{figure}
\centering
\psfrag{x}{\scriptsize{$\widehat{\beta T}_{vco}$} }
\psfrag{y}{\scriptsize{BER}}
\psfrag{data1}{\tiny{Method in \citep{Munier}, 5 DD-FB iteration, Free running VCO,}}
\psfrag{data2}{\tiny{Method in \citep{Munier}, 5 DD-FB iteration, PLL}}
\psfrag{data3}{\tiny{Proposed method, 2 DD-FB iteration, Free running VCO}}
\psfrag{data4}{\tiny{Proposed method, 2 DD-FB iterations, PLL}}
\psfrag{data5}{\tiny{Method in \citep{Corvaja}, Free running VCO}}
\psfrag{data6}{\tiny{Method in \citep{Corvaja}, PLL}}
\includegraphics[scale=0.38]{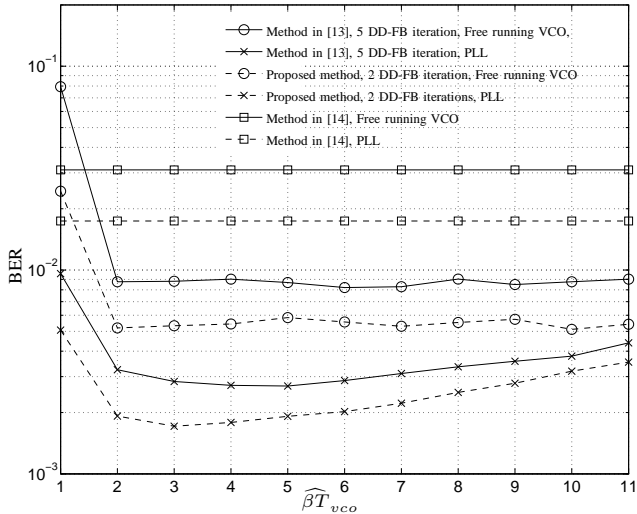}
\caption{Comparison of BER performance of the proposed algorithm and the method in \citep{Munier} in a free running VCO and PLL as a function of the input $\widehat{\beta T}_{vco}$ in which the true $(\beta T_{vco})=0.05$. A 16-QAM coded OFDM and a Rayleigh channel where $\SNR=20$dB used.}
\label{fig_pll}
\end{figure}

In Figure \ref{fig_pll}, the performances of the proposed method and the method in \citep{Munier} is compared for various input of the PHN variance of the VCO, which is denoted by $\widehat{\beta T}_{vco}$, for a free running VCO and for PLL at 20dB $\SNR$. Both methods are derived based on the Weiner PHN assumptions. The proposed method is run in 2 DD-FB loops while the method in \cite{Munier} is run in 5 DD-FB loops. The improved in performance for  PLL (dashed line) compared to free running VCO (solid line) can be seen in Figure \ref{fig_pll} for both the proposed as well as the method in \citep{Munier}. It can be seen that the lowest BER for the proposed method in a PLL is achieved around the input $\widehat{\beta T}_{vco}=0.02$ which then starts to increase as $\widehat{\beta T}_{vco}>0.02$. The increasing trend in BER can also be seen in free running VCO when $\widehat{\beta T}_{vco}>0.04$ for the proposed method. The BER performance of the method in \citep{Munier}, on the other hand, seems to be stable around $5\times 10^{-3}$ in a PLL and around $9\times 10^{-3}$ in a free running VCO for an input $\widehat{\beta T}_{vco}$ equal to 0.01 and above. For the range of input $\widehat{\beta T}_{vco}$ values displayed, the proposed method outperforms  \cite{Munier} in both cases when a PLL and a free running VCO is used.   

\section{Computational analysis \label{sec:6}}
The proposed scheme does incur some computations which are not part of the reference schemes \cite{Munier, Corvaja}. These include the multiplication with the matrix $\mathbf{\widehat{A}}_k^{-1}$ in \eqref{eq19}, and all the other calculations inside the
loop over $k$ in Algorithm \ref{alg2}. It should be noted that the channel estimation in \eqref{eq25}, \eqref{eq27} and \eqref{eq28} only requires the multiplication of the LS channel estimate with a pre-computed matrix. The multiplication with $\mathbf{\widehat{A}}_k^{-1}$ is an operation in the frequency domain. More efficient would be to do the codebook pre-compensation in the time domain and do a separate FFT for each of the $K$ possible received signals. In this case, the number of complex multiplications to obtain $\tilde{\bf Y}^k_m$ for $k=1,...,K$ is on the order of $N(1+\frac{K}{2}\log_2 N)$, rather than $N(KN +\frac{1}{2}\log _2N)$. A summary of the number of complex additions and multiplications required by the most important equations of our algorithm is given in Table \ref{tab3a}.  

\begin{table}
\centering
\caption{Summary of total Number of complex operations required per OFDM symbol and DD-FB iteration excluding the Viterbi and a few other computations. \label{tab3a}}
\begin{tabular}{llll}
Equation &  Additions  & Multiplications  & Complex  inversions\\ \hline
\eqref{eq19} &    $KN(N-1)$    & $KN^2$  & $-$  \\
\eqref{eq26} &   $-$  & $KN$  &  $ KN$ \\
\eqref{eq25} &   $KN(N-1)$ & $KN^2$   & $-$  \\
\eqref{eq24} &   $(K+1)(N-1)$   & $K(2N+1)+N$    & $KN+1$\\
\eqref{eq29} &   $-$    & $2KN$   & $K(N+1)$  \\
\eqref{eq23} &   $KN(N-1)$    & $KN$ & $-$   \\
\hline
\end{tabular}
\end{table}

\begin{table*}
\centering
\caption{Summary of total Number of complex operations required per OFDM symbol excluding the Viterbi and a few other computations. \label{tab5a}}
\begin{center}
\begin{tabular}{lll}
Method & No. of Additions  & No. of Multiplications  \\ \hline
Proposed method &    $(i+1)(KN(3N-2)-k+N-1)$    & $(i+1)(K(2N^2+6N+1)+N)$    \\
\citep{Munier} &   $i(\Lambda N(N-1) + N(N-2)(\Lambda-2)(\Lambda-1)/2) $ & $ i(\Lambda(4N^2+2N+1) +\Lambda(\Lambda-1)N(2N-1)/2) $    \\
& \ \ \quad $+(i+1)2N(N-1)$ & \ \ \quad $+(i+1)(2N(N+1))$\\
\hline
\end{tabular}
\end{center}

\end{table*}

\begin{figure}
\centering
\psfrag{x}{\scriptsize{Averaged Execution Time per Frame (sec)} }
\psfrag{y}{\scriptsize{BER}}
\psfrag{a}{\tiny{$0^*$}}
\psfrag{b}{\tiny{$1^*$}}
\psfrag{c}{\tiny{$2^*$} }
\psfrag{d}{\tiny{$3^*$}}
\psfrag{e}{\tiny{$4^*$}}
\psfrag{f}{\tiny{$5^*$} }
\psfrag{g}{\tiny{$6^*$} }
\psfrag{h}{\tiny{$7^*$}  }
\psfrag{i}{\tiny{$8^*$} }
\psfrag{j}{\tiny{$9^*$} }
\psfrag{k}{\tiny{$10^*$} }
\psfrag{data1}{\tiny{Method in \cite{Munier}}}
\psfrag{data2}{\tiny{Proposed method (Algorithm \ref{alg2}))}}
\psfrag{Data 3}{\tiny{No. of iterations}}

\includegraphics[scale=0.38]{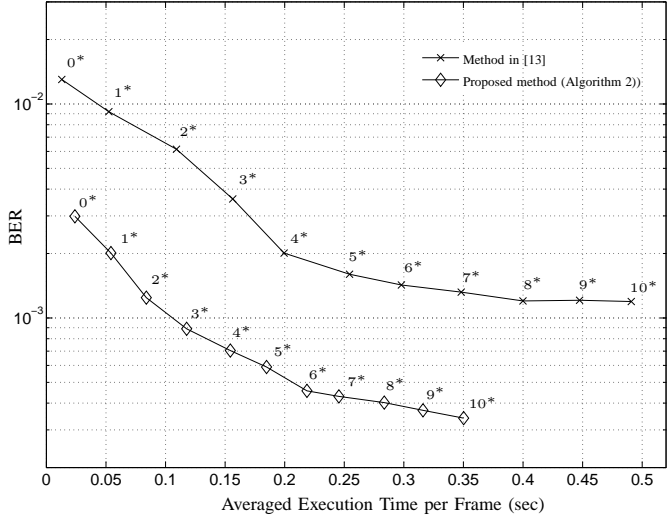}
\caption{BER performance of Algorithm \ref{alg2} and the algorithms in \cite{Munier} on 64-QAM OFDM as a function of averaged execution time  per frame (in sec) for $K=27$ and $\Lambda=6$ at $24$dB SNR  using various number of iterations (denoted by $i^*$) in the DD-FB loop.}
\label{fig9aa}
\end{figure}

The method in \cite{Munier} involves some computations which are not needed in our algorithm for instance the successive least squares estimation of the coefficients of the power series which approximates the PHN in addition to an $N \times N$ Toeplitz matrix inversion. The overall number of complex additions and multiplications  for channel estimation and PHN compensation required for one OFDM symbol by the proposed method as well as the method in \cite{Munier}  is given in Table \ref{tab5a}, excluding the Viterbi algorithm and without taking into account the $N \times N$ Toeplitz matrix inversion for \cite{Munier}.  The parameter $\Lambda$ in Table \ref{tab5a} refers to the order of the polynomial approximating the ICI coefficients as presented in \cite{Munier}. It can be seen that the complexity of both the proposed algorithm and the method in \cite{Munier} are linearly dependent on $K$ and $\Lambda$ respectively. For $K=27$, $\Lambda=6$
and $N=64$ the proposed algorithm has a computational advantage over \cite{Munier}. This is illustrated in  Figure \ref{fig9aa} which compares the measured execution time for various number of DD-FB iterations between the proposed algorithm (Algorithm \ref{alg2}) and the algorithms in \cite{Munier} for 64-QAM OFDM at 24dB SNR. No DD-FB loop or 0 iteration for the method in \cite{Munier} implies channel estimation with CPE compensation which reduces the number of computations significantly. The $N \times N$ Toeplitz matrix inversion in \cite{Munier} has been factored out of the measured execution time (since we did not take into account the Toeplitz structure). Our simulator is a Linux system (Ubuntu 11.04) running on a CPU with Intel-Core i7-2600 and uses the IT++ library in C++ for executing the algorithms. The results in Figure \ref{fig9aa} show that, for the chosen parameter $K=27$, the proposed method provides a significantly better performance-complexity trade-off than the method in \cite{Munier}. 

The DD-FB stage in \citep{Petrovic} and \citep{Munier} are not easily parallelized, while the most computationally demanding operations in the proposed  algorithm can obviously be split into up to  $K$ parallel processors or dedicated hardware. The algorithms in \cite{Petrovic} and \cite{Corvaja} also involve an inverstion of an $N \times N$ (Toeplitz) matrix. However, the method in \cite{Corvaja} also presents a solution in which only the ICI coefficients close to the carrier need to be estimated. This requires solving a less complex linear system, e.g. based on tridiagonal matrix algorithm which only requires $\mathcal{O}(N)$ complex operations. This makes \cite{Corvaja} less computationally complex to implement than the proposed method and \cite{Munier}. In addition [14] has the advantage provided by avoiding the requirement of having to know the channel and PHN statistics. These advantages come at the cost of performance in BER as shown in Figure \ref{fig7} and  Figure \ref{fig8}. 

\section{Conclusion\label{sec:7}}
Based on a simple codebook table which approximates the phase-noise statistics by a finite number of realizations, a novel PHN compensation approach is introduced. The general idea can be applied to a wide range of PHN scenarios.  Herein, we have concentrated on ICI suppression in a convolutional encoded OFDM system with scattered pilots. The complexity of the algorithm is determined by the size of the codebook. 

Using a codebook of moderate size, $K=27$, for a 16-QAM uncoded OFDM system in which an AWGN channel and a PHN with $BT=0.01$, which is   was considered, the proposed algorithm is 2dB from the ideal PHN free case at a BER of $10^{-3}$. Considering an identical scenario in a 64-QAM uncoded OFDM, the proposed algorithm is 2.5dB from the ideal PHN free case at a BER of $10^{-2}$. By comparison, the method in \citep{Munier} is $9$dB from the ideal case. 

For a coded system, the codebook technique is used along with DD-FB loop for combined channel estimation and PHN compensation. For a 16-QAM OFDM system over a Rayleigh fading channel with $BT=0.01$, the proposed method is 1.5dB from the ideal case (PHN free and known channel) at a BER of $10^{-4}$ employing 2 iterations on the DD-FB loop; while the method in \cite{Munier} is $2.5$dB from the ideal case using 8 iterations. For a 64-QAM coded system and PHN with $BT=0.005$, the proposed method provides a $1$dB gain at a BER $10^{-3}$ employing 3 iterations in the DD-FB, compared to the method in \citep{Munier} which uses 10 iterations. The gain obtained by the proposed method is significant compared to the method in \cite{Corvaja}, which uses a pilot interpolation technique for channel estimation.  Additional results are also presented for both 16 and 64 QAM coded systems assuming known Rayleigh channel frequency response at the receiver. The proposed method is shown to outperform to the reference schemes without having to employ the DD-FB loop.

The performance of the proposed method with respect to the codebook size is also evaluated showing the relative improvement in BER as $K$ increases. However, for the given example, the relative improvement is very small as $K$ increases above 27. Simulation results are also shown employing PHN from a charge pump PLL, modelled as Ornstein-Uhlenbeck process, is also presented with the proposed method showing improved performance compared to the methods in \citep{Munier} and also demonstrating the applicability of the proposed algorithm for a PHN process other than Weiner process. 

Therefore, with a moderately small codebook size and limited number of iterations on DD-FB, an improved performance as well as faster execution time is achieved by the proposed algorithm. 

\appendices
\section{Proof of Equation \eqref{eq17}  \label{sec:app}}
Let us then first consider the case when the number of quantization regions, $Q$, is infinite. In such a case, each segment of the selected codebook entry will follow the mean of samples of the corresponding segment of the PHN realization. Thus the MSE for infinite $Q$ is given by
\begin{equation}\label{eq16app}
\begin{split}
\mbox{MSE}_{(Q=\infty)}&=\sum_{j=0}^{J-1}\sum_{n=0}^{L-1} E_{\theta(n)}[(\theta(n+jL) - \hat{\theta}(n+jL))^2]\\
& =\sum_{j=0}^{J-1}\sum_{n=0}^{L-1} E[(\theta(n+jL)^2 \\ 
&   -2\theta(n+jL)\hat{\theta}(n+jL)  +\hat{\theta}(n+jL)^2]\\
 & =\sum_{j=0}^{J-1}\sum_{n=0}^{L-1} E[(\theta(n+jL)^2\\
 &  - \frac{1}{L}\sum_{j=0}^{J-1}\sum_{n=0}^{L-1}\sum_{l=0}^{L-1}E[\theta(n+jL)\theta(l+jL)]\\
 &=\frac{(N-J)(N+J)}{6J}\sigma_\epsilon^2
\end{split}  
\end{equation}
in which the last equality follows from the covariance of a random walk process which is  $E[\theta(n+\ell)\theta(n)]=n\sigma_\epsilon^2$ for $l\geq0$. 

When $Q$ is finite, an additional error is introduced due to the quantization of segment averages. As in the design of the codebook, we assume that $\psi$ is set to the average of the first segment and thus no quantization error occurs in the first segment and the PHN realization crosses the first sample of each segment of the trajectory. Additionally, by assuming that this error in each segment is identically distributed and independent of the difference between the PHN and the codebook trajectory, the MSE due to accumulated quantization error in each segment is $L\sigma_q^2$, where $L=N/J$ and $\sigma_q$ is the standard deviation of quantization error. The MSE due to the quantization error over the entire trajectory is therefore, $\MSE_{(Q<\infty)}=(J-1)L\sigma_q^2$, where $(J-1)$ is due to the assumption that the first segment has no quantization error. The approximate MSE of the selected trajectory as an estimator of the PHN realization then becomes
\begin{equation}\label{eq17app}
\mbox{MSE} =\frac{(N+J)(N-J)}{6J}\sigma_\epsilon^2 + L(J-1)\sigma_q^2.
\end{equation}
In order to determine $\sigma_q^2$, we assume that the PHN realization is exactly on one of the codebook entries in the last sample of the previous segment. In this case, the quantization error of the $j$-th segment average is equal to the quantization error of the $j$-th increment, i.e., no influence of previous increments. The variance, $\sigma_q^2$, of the quantization error  is then given by 
\begin{equation}\label{eq17n}
\begin{split}
\sigma_q^2 &= \sum_{i=1}^Q E[(X-\hat{x})^2|X\in \mathcal{R}_i] \mathcal{P}(X\in \mathcal{R}_i))\\ 
&= \sum_{i=1}^Q \mathcal{P}(X\in \mathcal{R}_i)(E[X^2|X \in \mathcal{R}_i] -2\hat{x}_iE[X|X \in \mathcal{R}_i] +  \hat{x}_i^2)\\
&= \sum_{i=1}^Q \mathcal{P}(X\in \mathcal{R}_i)(E[X^2|X \in \mathcal{R}_i] -  \hat{x}_i^2)\\
\end{split}
\end{equation}
where the last equality follows from $E[X|X\in\mathcal{R}_i]=\hat{x}_i$ which is given by \eqref{eq15}. The conditional Expectation $E[X^2|X \in \mathcal{R}_i]$ is given by
\begin{equation}\label{eq14app}
\begin{split}
E\bigg[X^2|X\in R_i\bigg]&=\frac{\int_{R_i}x^2f_X(x)dx}{\mathcal{P}(X\in \mathcal{R}_i)}\\
\end{split}
\end{equation}
where $\mathcal{P}(X\in \mathcal{R}_i)=\int_{R_i}f_X(x)dx$. When $f_X(x)$ a Gaussian function 
\begin{equation}\label{eq15aa}
\begin{split}
\int_{R_i}&x^2f_X(x)dx =\sigma_x^2/2\bigg(\erf(x_{i+1}/\sqrt{2}\sigma_x)-\erf(x_{i}/\sqrt{2}\sigma_x)\bigg)\\
& - \sigma_x\sqrt{2\pi}\bigg(x_{i+1}\exp(-0.5x_{i+1}^2/\sigma_x^2)-x_i\exp(-0.5x_{i}^2/\sigma_x^2)\bigg)
\end{split}
\end{equation}
and 
\begin{equation}\label{eq15bb}
\begin{split}
\mathcal{P}(X\in \mathcal{R}_i)&=\int_{R_i}f_X(x)dx\\
&=0.5 \bigg(  \erf(x_{i+1}/\sqrt{2}\sigma_x)-\erf(x_{i}/\sqrt{2}\sigma_x) \bigg).
\end{split}
\end{equation}
Equations \eqref{eq15aa}, \ref{eq15bb} and \eqref{eq15} are then used on \eqref{eq17n} to obtain the variance, $\sigma_q^2$, of the quantization error.

\ifCLASSOPTIONcaptionsoff
  \newpage
\fi

\bibliographystyle{IEEEtran}
{\footnotesize
\bibliography{IEEEabrv,references}}
\vspace{-170 mm}

\end{document}